\documentclass[letterpaper,twocolumn,10pt]{article}
\usepackage{usenix2019_v3}

\usepackage{tikz}
\usepackage{amsmath}
\usepackage[small,compact]{titlesec}

\usepackage{filecontents}
\usepackage{xspace}
\usepackage{booktabs} %
\usepackage{makecell}
\usepackage{url}
\usepackage{subfig}
\usepackage{ulem} %
\usepackage[export]{adjustbox} %
\usepackage{multirow}
\usepackage{enumitem}
\usepackage{placeins}

\newif\ifdraft
\drafttrue %

\renewcommand{\emph}[1]{\textit{#1}}

\newcommand{\sys}{PolyServe\xspace}
\newcommand{\pdsys}{PD-\sys{}}
\newcommand{\cosys}{CO-\sys{}}

\newcommand{\pd}{PD-Disaggregate}
\newcommand{\co}{Co-location}

\newcommand{\fig}{Figure~}

\newcommand{\datagoodputgainpd}{$1.23\times$\xspace}
\newcommand{\datagoodputgainco}{$1.18\times$\xspace}
\newcommand{\datagoodputpdopt}{$92.5\%$\xspace}
\newcommand{\datagoodputcoopt}{$72.9\%$\xspace}
\newcommand{\databurstgainpd}{$1.33\times$\xspace}
\newcommand{\databurstgainco}{$1.36\times$\xspace}

\newcommand{\gpus}{$\mathrm{instance} \cdot \mathrm{second}$}

\begin{document}

\date{} 

\title{\Large \bf \sys{}: Efficient Multi-SLO Serving at Scale}
\author{
{\rm Your N.\ Here}\\
Your Institution
\and
{\rm Second Name}\\
Second Institution
} %

\author{
{\rm Kan Zhu}\\
University of Washington
\and
{\rm Haiyang Shi}\\
ByteDance
\and
{\rm Le Xu}\\
ByteDance
\and
{\rm Jiaxin Shan}\\
ByteDance
\and
{\rm Arvind Krishnamurthy }\\
University of Washington
\and
{\rm Baris Kasikci}\\
University of Washington
\and
{\rm Liguang Xie}\\
ByteDance
}

\maketitle

\begin{abstract}
Advances in Large Language Models (LLMs) have led to a surge of LLM‑powered applications. These applications have diverse token‑generation latency requirements. As a result, simply classifying workloads as latency‑sensitive (LS) or best‑effort (BE) overlooks the nuances within the latency‑sensitive category and results in suboptimal user experiences and scheduling opportunities. However, efficiently serving requests with multiple SLO requirements poses significant challenges. First, all requests within a batch generate new tokens simultaneously, which can misalign them with their distinct time‑per‑output‑token requirements. Moreover, while existing systems focus on auto‑scaling for handling various overall request rates, the diversity of SLOs necessitates fine‑grained auto‑scaling among these SLO tiers. Finally, unlike LS/BE scenarios—where BE requests can be aborted at any time to ensure the SLO attainment of LS requests—those with different latency‑sensitive SLOs cannot tolerate prolonged delays, and tail latency must be controlled.

To tackle these challenges, we propose \sys{}, a novel multi‑SLO scheduling policy at scale that maintains high SLO attainment while maximizing throughput. \sys{} first groups requests into multiple bins based on their per‑token latency requirement, then schedules each bin to a subset of the server fleet. \sys{} routes requests to the highest‑load but still SLO‑attainable server to create a load gradient that facilitates auto‑scaling. To increase utilization, \sys{} permits looser‑SLO requests to share tighter‑SLO instances when their own servers are saturated. \sys{} uses profiling data to guide scheduling decisions and manage tail latency through request‑wait‑time‑aware scheduling, dynamic chunking, and continuous chunked prefill prediction. \sys{} achieves \datagoodputgainpd{} goodput gain compared to existing policies, achieving up to \datagoodputpdopt of optimal goodput.

\end{abstract}

\section{Introduction}
\label{sec:intro}

Large language models are strongly integrated into work and life, powering various applications such as online chat, code completion, real‑time question answering, code review, agents, and document summarization~\cite{Mehdi_2023, Spataro_2023, ChatGPT}. To ensure a good user experience, various tasks have distinct latency requirements, which are regulated by Service Level Objectives (SLOs), defined as time to first token (TTFT) and time per output token (TPOT)~\cite{sarathi-serve, chen2025slosserveoptimizedservingmultislo}.

Simply categorizing workloads as either “best‑effort” or “latency‑sensitive” (LS/BE) based on the SLO does not fully capture their diverse nature. For instance, multi‑step agent systems are extremely latency‑sensitive, as the user must wait for the output of the last agent~\cite{rawat2025preactmultistepplanningreasoning}. For online chat, generating text at human reading speed suffices~\cite{distserve}. Code review typically has more relaxed latency requirements, though users still expect results within about one minute~\cite{lin2025codereviewqacodereviewcomprehension}. Meanwhile, workloads like document summarization and synthetic data generation truly fall under the best‑effort category. Moreover, users have different willingness to pay for various SLO levels: for example, students are more price‑sensitive, while companies prioritize user experience. Therefore, extending LS/BE to multi‑SLO serving is economically necessary.

This pricing strategy also aligns with the serving cost. In LLM serving, General Matrix Multiplications (GEMMs) with model weights exhibit a significant batching effect, as large batches can amortize the weight loading cost~\cite{vllm}. For service providers, a tighter SLO restricts request batching, leading to lower hardware utilization and higher costs.

While offering arbitrary SLO levels gives users flexibility, it greatly complicates system design. To simplify, we assume the provider offers fixed SLO choices indicated by (TTFT, TPOT) pairs. For example, when serving small models, the provider can offer (1000\,ms, 20\,ms), (1000\,ms, 30\,ms), (2000\,ms, 50\,ms), and best‑effort (12\,h, 12\,h).

\begin{figure*}
    \centering
    \includegraphics[width=0.9\linewidth]{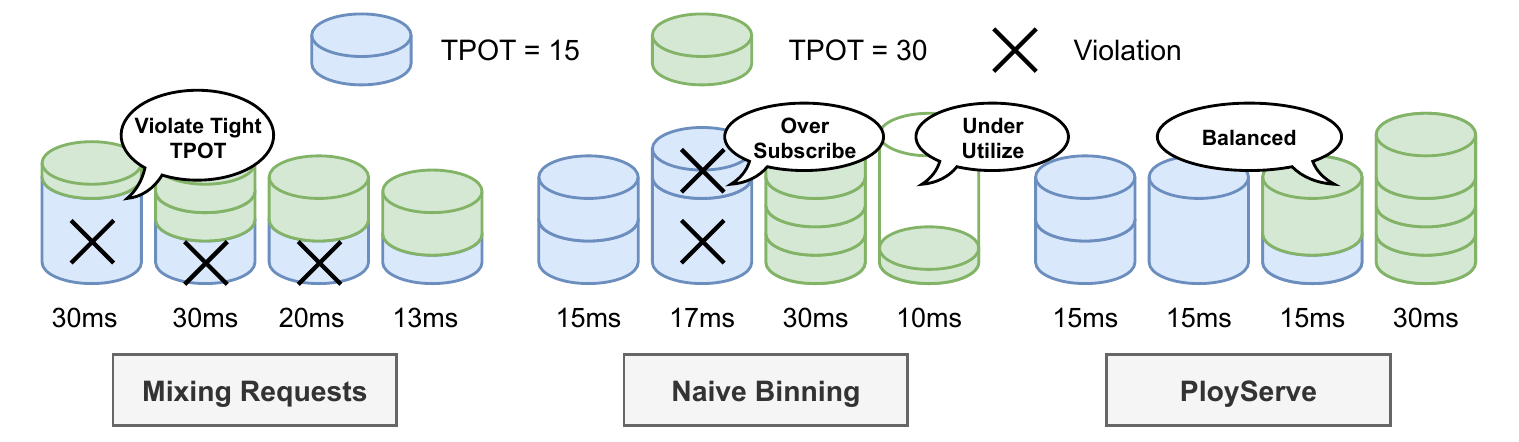}
    \caption{Comparison between \sys{} and existing systems. When mixing requests with different SLOs, tighter‑SLO requests miss their deadlines. Simple binning also suffers from imbalanced workloads. \sys{} achieves balanced, high utilization across the cluster.}
    \label{fig:teasor}
\end{figure*}

However, serving fixed SLO choices remains challenging. First, serving engines generate new tokens for batched requests simultaneously~\cite{orca}. When handling requests with (1000\,ms, 20\,ms) and (1000\,ms, 30\,ms) on the same instance, the server’s iteration time must be controlled to 20\,ms, which is overkill for 30\,ms TPOT requests.

Additionally, unlike LS/BE—where auto‑scaling only applies to the LS cluster—both the popularity of different SLOs and the overall request arrival rate can vary over time. Thus, auto‑scaling becomes more frequent, incurring higher overhead since LLM serving is stateful and removing a server requires emptying it. 

Finally, compared to LS/BE co‑scheduling, we cannot sacrifice SLO attainment for any requests; tail latency management is essential for ensuring high SLO attainment rates.

To this end, we propose \sys{}, a novel routing policy for high‑throughput, multi‑SLO serving that supports both prefill‑decode disaggregation and co‑location (chunked prefill) at large‑scale deployment. We illustrate this comparison in Figure~\ref{fig:teasor}.

First, \sys{} observes that optimal serving throughput is achieved when every request is scheduled to the cluster operating at the largest possible batch size limited by SLO or GPU memory. Scheduling requests with different TPOTs on the same server is suboptimal due to unnecessary resource spending on lower‑tier requests. Therefore, \sys{} classifies requests based on their TPOT requirements and partitions servers into clusters, each serving a single TPOT tier to minimize mixing.

To manage cluster size per SLO tier, \sys{} uses fine‑grained auto‑scaling. Specifically, when the request rate exceeds current capacity, \sys{} allows looser‑SLO requests to fill available slots on tighter‑SLO instances. This promotion is lazy, occurring only when the lower tier is full, reducing auto‑scaling frequency and increasing throughput. To facilitate scaling down as rates drop, \sys{} always schedules each request to the highest‑load machine that can still meet its SLO, building a load gradient that makes the last server easier to remove.

This aggressive scheduling approach eases auto‑scaling but makes tail latency management critical, as any variation can cause an SLO violation. Therefore, instead of adhering strictly to user‑level SLOs (TTFT and TPOT), \sys{} uses deadline‑based SLOs~\cite{wang2024revisitingslogoodputmetrics}, i.e., the $i$th token must be generated by $\mathrm{TTFT} + i\times\mathrm{TPOT}$, to provide the same experience while enabling more efficient scheduling. When scheduling within each tier, \sys{} predicts the serving engine’s iteration time based on profiling data that maps batch size and the KV cache size to execution time. \sys{} admits requests only if the predicted iteration time is below the TPOT, with prediction errors absorbed by the deadline‑based SLO.

Moreover, to further control tail latency, \sys{} considers each request’s wait time—the time spent waiting for the current batch to finish—and adds that to the predicted iteration time for a more accurate SLO estimate. Additionally, \sys{} manages TTFT through dynamic chunking for prefill‑decode disaggregation, merging the last chunk of the prefill request into the prior iteration. For co‑location, \sys{} predicts chunk sizes throughout prefill so that even if the decode sequence grows, the prefill finishes before TTFT.

We implement \sys{} in C++ for high performance and evaluate it using a simulator for large‑scale deployment. \sys{} achieves a \datagoodputgainpd{} and \datagoodputgainco{} throughput gain for prefill‑decode disaggregated and co‑located systems, respectively, on real‑world datasets at various SLO choices, reaching \datagoodputpdopt{} and \datagoodputcoopt{} of optimal throughput.

In summary, we contribute the following:
\begin{itemize}[noitemsep, topsep=0pt]
  \item A detailed analysis of the necessity of SLO‑tiered pricing and the challenges of serving multi‑SLO workloads.
  \item \sys{}, a high‑throughput multi‑SLO serving system at scale that classifies requests by TPOT and manages auto‑scaling per SLO tier.
  \item A comprehensive evaluation showing \sys{} achieves \datagoodputgainpd{} and \datagoodputgainco{} throughput gains over existing systems.
\end{itemize}

\section{Background}

\subsection{LLM Inference Workflow}
\label{sec:bg:inference}
Inference of transformer-based large language models consists of a prefill phase and a decode phase~\cite{sarathi}. In the prefill phase, the prompt goes through the model in one pass to generate the first output token and populate the KV cache for future generations. In each decode phase, the previous token is fed into the model to autoregressively generate one additional token. The decoding process continues until it reaches a special end-of-sequence (EOS) token or the maximum decode length.

Zooming in on the transformer model, it comprises several identical transformer layers. In each layer, the input passes through an attention module, which computes the key (K), query (Q), and value (V) vectors from the input embedding and performs self-attention, i.e., $\mathrm{Softmax}\bigl(\tfrac{QK^{T}}{\sqrt{d}}\bigr)V$, using the K and V vectors of all previous tokens. The output then goes through an O‑projection and enters the feed‑forward network, which includes an up projection, a gate projection, an activation function, and a down projection. We apply RMS normalization and add a residual connection before each module.

When a model exceeds a single GPU’s capacity, it is distributed across multiple GPUs or servers via tensor parallelism or pipeline parallelism~\cite{alpaserve}. We define the smallest GPU cluster serving a model as a serving instance.

\subsection{Operational Characteristics}
Transformer operations exhibit a batching effect, meaning that larger input batch sizes generally yield better performance. Some operations, such as key, query, and value vector generation, output projection, and the up, gate, and down projections, involve matrix multiplications (GEMMs) with model weights. Increasing the batch size amortizes the cost of loading weights, thereby significantly improving efficiency~\cite{zhu2025nanoflowoptimallargelanguage}. Other operations, such as decode-attention, do not benefit significantly from batching, as different requests have distinct histories and cannot share KV cache loading~\cite{tang2024quest}. Therefore, the efficiency of LLM serving is largely determined by the GEMM batch size.

In the prefill phase, a single request can provide thousands of tokens, effectively saturating GEMM performance. In contrast, during the decode phase, each request generates only one token. Even when requests are batched, it is difficult to achieve the bandwidth required to saturate GEMM performance.

\subsection{LLM Serving SLOs}
To ensure a good user experience, service providers must attain the Service Level Objectives (SLOs) defined by user requirements. Common metrics used to define SLOs include Time to First Token (TTFT), measured from request submission to the generation of the first token, and Time per Output Token (TPOT), measuring the time between consecutive tokens. While some works target average TPOT, service providers may delay output tokens and return them all at once, which can create noticeable stalls for users. In contrast, adhering to maximum TPOT can be too restrictive to allow flexible scheduling. To maintain user experience while allowing greater scheduling flexibility, prior works have proposed a deadline-based SLO (DSLO)~\cite{wang2024revisitingslogoodputmetrics}. Specifically, the $i$-th token must be produced before $\mathrm{TTFT} + i\cdot\mathrm{TPOT}$. When this DSLO is satisfied, providers can delay the delivery of the $i$-th token until $\mathrm{TTFT} + i\cdot\mathrm{TPOT}$ while still meeting strict TTFT + TPOT SLOs. Before the deadline, providers may schedule token generation freely. Due to this flexibility, we adopt DSLO in this paper.

\subsection{Prefill-Decode Disaggregation and Chunked Prefill}
Previous studies have examined strategies to improve SLO attainment, particularly by addressing interference between the prefill and decode phases. With naive continuous batching, admitting a long prefill request can cause decode requests to violate their SLOs, as the prefill phase extends iteration time. Two distinct approaches have been proposed: prefill-decode disaggregation (\pd{})~\cite{distserve} and chunked prefill (\co{})~\cite{sarathi-serve}.

In prefill-decode disaggregation, servers are partitioned into prefill and decode clusters. The prefill phase of a request is scheduled on the prefill cluster. Once complete, the KV cache is transferred to the decode cluster to continue token generation, eliminating interference and allowing each cluster to scale independently.

In chunked prefill, the prefill and decode phases are co-located on a single serving instance. In each iteration, the total GEMM input batch size, comprising both prefill and decode tokens, is limited. Decode requests are prioritized, i.e., all current decode requests are scheduled in the next iteration, and remaining capacity is filled with a chunk of prefill tokens. This limits prefill-decode interference and improves SLO attainment.

Because KV cache sizes are large, implementing prefill-decode disaggregation demands high-speed networking infrastructure, ideally RDMA, for KV cache transfer. Thus, rather than favoring one approach, we offer solutions for both and compare their performance under varying workloads, enabling flexible adoption based on infrastructure capabilities and workload demands.

\section{Analysis}
\label{sec:analysis}

\subsection{SLO Requirements Vary Widely Across Workloads}
\label{sec:analysis_various_workloads}

LLM serving workloads are highly heterogeneous in terms of request complexity, token length, and interactivity. A simplistic division into “latency-sensitive” and “best-effort” workloads fails to capture this rich diversity.

For example, a typical chatbot response may require only a few hundred tokens, and a token-per-output-time (TPOT) budget of 100 ms is sufficient to maintain a responsive and fluid user experience~\cite{zhu2025nanoflowoptimallargelanguage}. In contrast, reasoning-intensive tasks, such as solving math word problems or performing multi-hop question answering, often spend thousands of tokens to articulate a detailed chain of thought before delivering a final answer~\cite{wei2023chainofthoughtpromptingelicitsreasoning}. If such tasks are placed in the same "latency-sensitive" tier as chat applications, and are constrained to the same TPOT budget (e.g., 100 ms/token), they can take several minutes to complete. This results in a sluggish user experience, despite formally meeting the per-token latency requirement. 

Similarly, agent-based systems, such as those powered by ~\cite{yang2023autogptonlinedecisionmaking} or LangChain~\cite{wei2023chainofthoughtpromptingelicitsreasoning}, initiate long chains of LLM prompts interleaved with API calls and tool invocations. These pipelines are complex and time-consuming, but users typically care only about the final output. Serving such systems in the same SLO tier may result in excessive waiting time.

Furthermore, use cases like document summarization, report generation, legal contract analysis, and code review typically tolerate moderate delays. Users expect a short wait and are willing to accept latencies on the order of seconds or minutes, making looser SLOs acceptable. However, for truly offline tasks, such as batch dataset labeling, summarizing large paper corpora, or periodically crawling and analyzing web data, users anticipate long delays (even hours) and can continue other work in the meantime. In short, LLM workloads span a broad spectrum: from interactive, low-latency sessions to long-form, throughput-bound jobs with soft deadlines. A coarse-grained classification fails to reflect this range, resulting in inefficient resource allocation, increased queuing delays, and suboptimal user experience.

\subsection{User Willingness to Pay Also Varies by Context}
Beyond workload differences, users themselves exhibit diverse latency expectations and willingness to pay even for the same task type. Consider two users both using a summarization API: a legal analyst preparing for a hearing may be willing to pay a premium for low latency to accelerate their workflow, whereas a student summarizing lecture notes overnight might prefer a cheaper option, accepting slower service. 

Similar diversity exists in real-time applications. A social media company might integrate an LLM into their platform for live user engagement and thus require strict SLO guarantees, while a research team testing new prompt formats for the same model may be content with delayed responses. Some startups may prioritize cost over speed in early prototyping stages but later require faster serving as they scale.

\subsection{SLO-Tiered Pricing as a Practical and Economic Necessity}
Despite this heterogeneity, most existing LLM providers segment users only by monthly usage limits, offering a single latency tier within each quota. As a result, users needing faster responses often have no option but to switch providers. This limitation forces providers into compromises that either overcharge low-budget users or fail to deliver reliable service to high-priority workloads. Some providers allow users to submit batched requests with loosened SLO constraints (e.g., 24h) at half of the price\cite{openai_pricing_2025, openai_batch_guide_2025}, but still not covering the whole spectrum. 

A more sustainable and scalable approach is to introduce multiple service tiers aligned with distinct SLO guarantees. For example, a “basic” tier could serve low-cost, loose-SLO tasks with background scheduling; an “advanced” tier could target moderate workloads with tighter performance constraints; and a “premium” tier could offer guaranteed latency for real-time or mission-critical applications. Furthermore, SLO-tiered pricing creates a natural upgrade path as user demands evolve. A research team may start with the basic tier, but migrate to advanced or premium tiers as they build commercial applications. This flexibility not only improves user satisfaction and retention but also expands the provider’s total addressable market. In short, fine-grained service differentiation based on SLO and pricing is essential for achieving high throughput, predictable quality of service, and long-term economic sustainability in LLM serving.

SLO-tiered pricing also aligns well with the request serving cost. We define the cost as \gpus, meaning the total time spent on the instance for serving this request. For example, with a batch size $B$ that is served on an instance of $N$ GPUs, taking $T$ to finish, the total cost is $NT$, and thus the per-request cost is $NT/B$.
\begin{figure}[t]
  \centering
  \includegraphics[width=\linewidth]{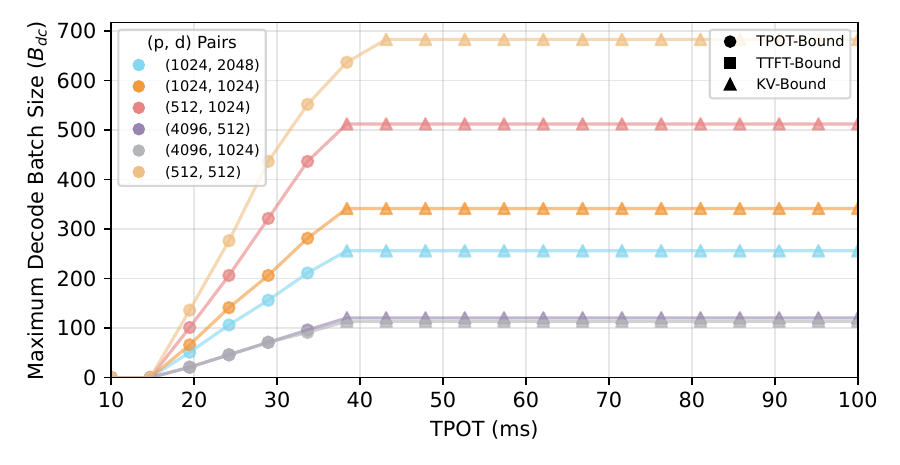}
  \caption{Relationship between decode batch size ($B_{\text{dc}}$) and per-token latency budget (TPOT).}
  \label{fig:bdc_vs_tpot}
\end{figure}

Given a pool of requests, the total workload for GEMM, prefill attention, decode attention, and other operations is fixed. Therefore, to reduce the cost, making every operation more efficient is the key. Specifically, prefill attention and decode attention are not sensitive to batching and have consistent performance. Therefore, the cost of serving is mainly determined by the batch size of GEMM, which have a significant batching effect. Tighter SLO, intuitively, would result in a limitation of batching requests, thus increasing the serving cost, justifying the higher price. 
\subsection{Deriving Batch Size Limits}
To quantitatively derive the limits of batch size, we need to consider both the constraints from GPU memory capacity and the SLO. We particularly consider serving a request with prefill length \(p\), decode length \(d\), GPU available KV cache token limit \(C\), per-token decode attention time \(t\), as well as corresponding TTFT and TPOT.

\begin{figure*}[t]
    \centering
    \includegraphics[width=1\textwidth]{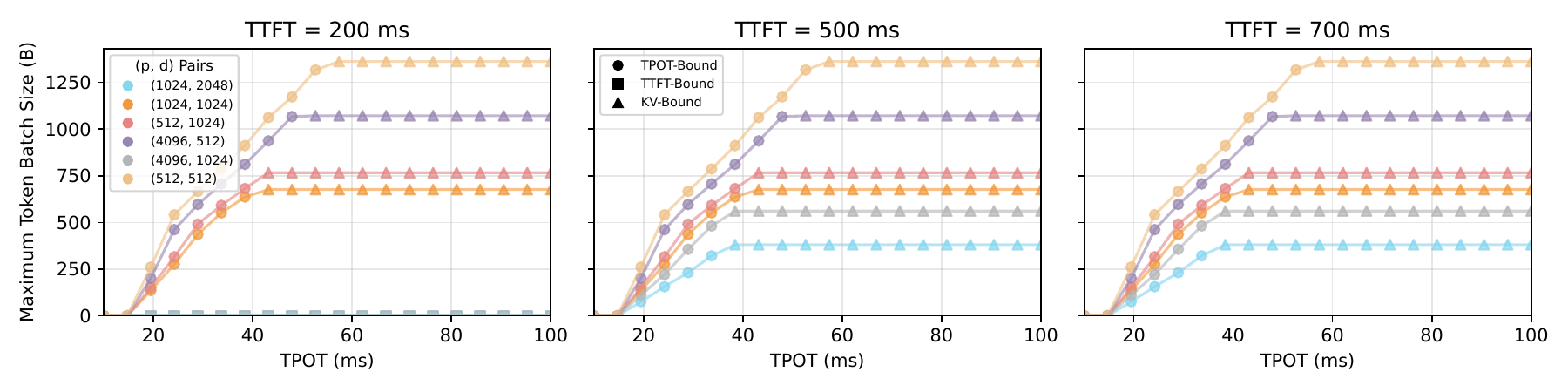}
    \caption{Maximum token batch size (B) as a function of Time Per Output Token (TPOT) for different prefill-decode configurations \((p, d)\) and Time To First Token (TTFT) budgets.}
    \label{fig:batch_size_vs_tpot}
\end{figure*}

\textbf{Prefill Decode Disaggregation.} For \pd{}, the prefill phase and decode phase are isolated. Therefore, the batch size of the prefill cluster is bounded by the TTFT and the GPU memory. Since the TTFT and GPU memory are usually large, the prefill batch size can easily reach 2048, nearly saturating the hardware. For the decode phase, the batch size \(B_{dc}\) must satisfy:
$$
GEMM(B_{dc}) + DcAttn(B_{dc}\times (p + d/2)) < TPOT,
$$
$$
B_{dc} \times (p + d/2) < C.
$$
We illustrate this in \fig\ref{fig:bdc_vs_tpot}.

\textbf{Co-location.}
When co-locating the prefill and decode phases and using a chunked prefill, the two phases interfere with each other, complicating the problem. From the decode perspective, the iteration time must be kept below the TPOT, which consists of GEMM time, decode attention time, prefill attention time, and other small operations. Given a token batch size of \(B\), the tokens for decode and prefill requests follow a \(d\!:\!p\) ratio. Therefore,
\[
B_{dc} = \frac{d}{p + d}\,B,
\qquad
B_{pf} = \frac{p}{p + d}\,B.
\]
The iteration time can be expressed as
\[
T_{\mathrm{iter}}
= \mathrm{GEMM}(B)
+ \mathrm{DcAttn}\!\Bigl(\tfrac{d}{p + d}\,B\,(p + \tfrac{d}{2})\Bigr)
+ \mathrm{PF}\!\Bigl(\tfrac{d}{p + d}\,B\Bigr).
\]
Since prefill attention, although compute-bound, typically uses a relatively small batch size \(B_{pf}\), its execution time is comparable to decode attention with the same existing KV‐cache length. Hence, we simplify:
\[
T_{\mathrm{iter}}
\approx \mathrm{GEMM}(B)
+ \mathrm{DcAttn}\!\Bigl(\tfrac{d}{p + d}\,B\,(p + \tfrac{d}{2}) + p\Bigr).
\]
Thus, from the decode perspective, the iteration time must satisfy:
\[
T_{\mathrm{iter}} < \mathrm{TPOT}.
\]

From the prefill perspective, the number of iterations needed to complete a chunked prefill is:
\[
N_{\mathrm{iter}}
= \frac{p}{\frac{p}{p + d}\,B}
= \frac{p + d}{B}.
\]
Therefore, to achieve the target time-to-first-token (TTFT), we require:
\[
N_{\mathrm{iter}} \cdot T_{\mathrm{iter}} < \mathrm{TTFT}.
\]

Finally, to ensure the KV cache fits within GPU memory, the system must obey:
\[
\frac{d}{p + d}\,B\,(p + \tfrac{d}{2}) + p < C.
\]

Again, we demonstrate the relationship between TTFT, TPOT, \(C\), and the token batch size \(B\) in \fig\ref{fig:batch_size_vs_tpot}.

\subsection{Calculating Optimal Cost}
Given the maximum batch size for each request, we can then calculate the minimal cost to serve that request. Particularly, for \pd{}, the cost is:
\[
\begin{aligned}
\text{Cost} ={}&\; p\frac{GEMM(B_{pf})}{B_{pf}} + \text{PF}(p) \\
& + d\frac{GEMM(B_{dc})}{B_{dc}} + DcAttn\Bigl(d(p + \tfrac{d}{2})\Bigr)
\end{aligned}
\]

For \co{}, the cost is:
\[
(p+d)\frac{GEMM(B)}{B} + \text{PF}(p) + DcAttn\Bigl(d(p+d/2)\Bigr)
\]

\begin{figure}[t]
    \centering
    \includegraphics[width=\columnwidth]{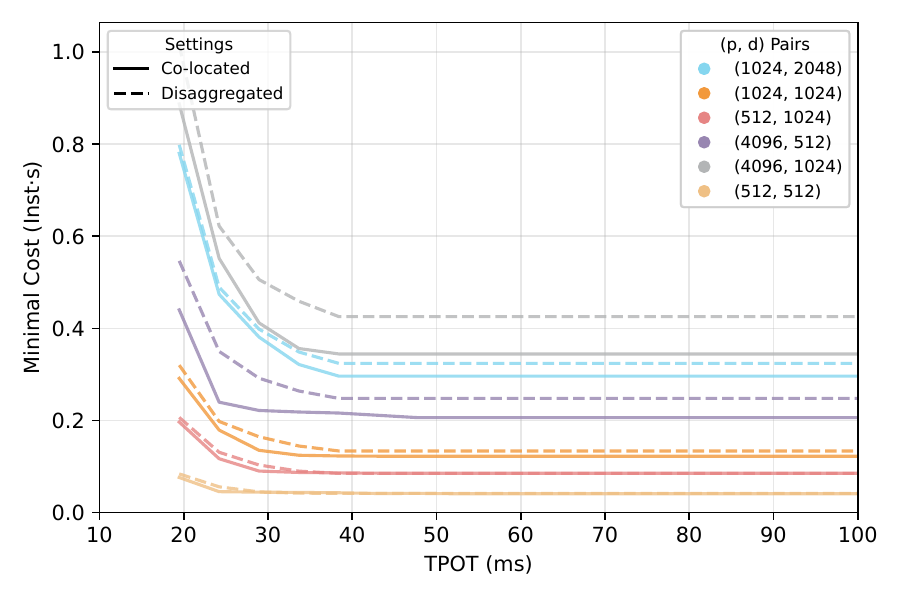}
    \caption{Comparison of system serving cost versus TPOT for different \((p, d)\) configurations under a 700~ms Time To First Token (TTFT) budget. Solid lines represent the co-located architecture, while dashed lines represent the disaggregated architecture.}
    \label{fig:cost_comparison}
\end{figure}

We illustrate the cost difference across varying \(p\), \(d\), TPOT, and serving methods in \fig\ref{fig:cost_comparison}. For short sequences, \co{} and \pd{} do not incur a large difference, while for long sequences, \co{} features lower cost.

\begin{figure*}[t]
    \centering
    \includegraphics[width=\textwidth]{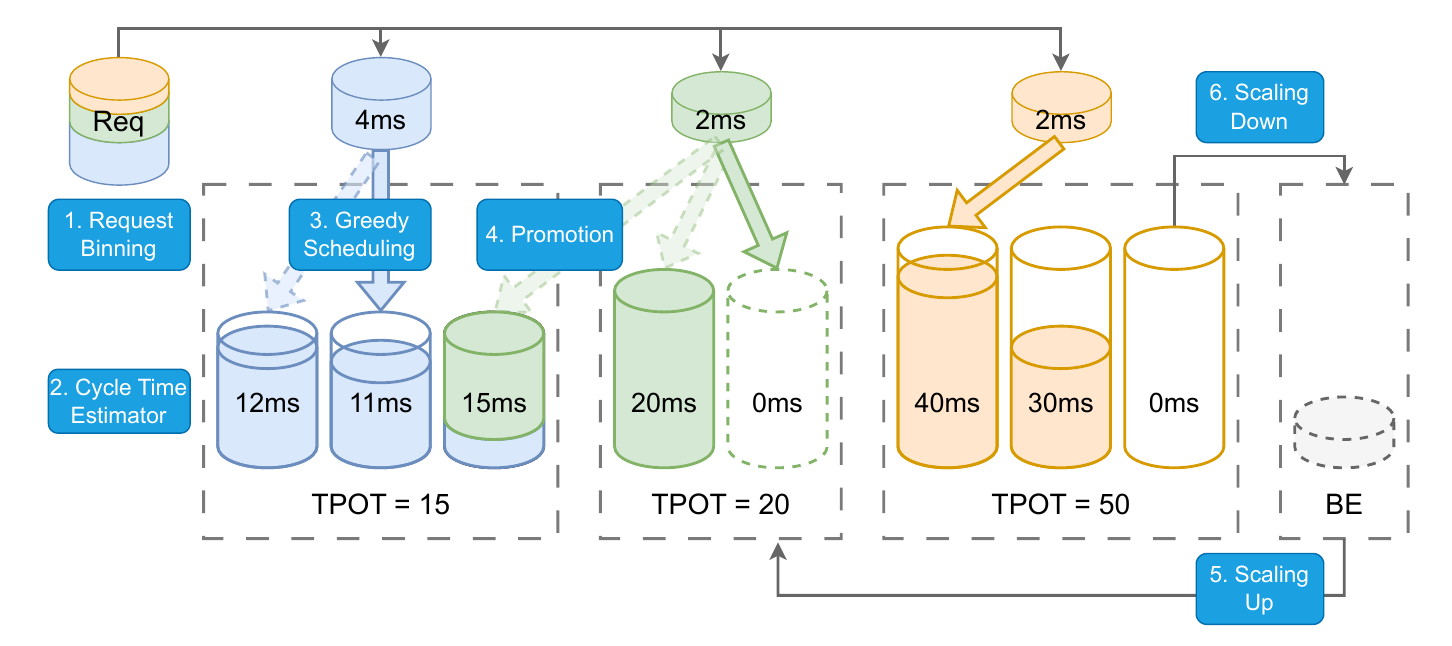}
    \caption{Overview of \sys{}. The cluster is partitioned into groups, each responsible for handling requests associated with a specific TPOT. Based on cycle time estimations, the scheduler greedily assigns requests to the most heavily loaded server that can still meet the SLO. To improve utilization, requests with looser SLOs are temporarily allowed to occupy servers dedicated to tighter SLOs, deferring auto-scaling when possible. Due to the greedy assignment, when the last server is empty, the down scaling is triggered.}
    \label{fig:slo_main}
\end{figure*}
\subsection{Serving Multi-SLO Requests is Challenging}
While forming the largest possible batch size for each request is essential to achieving higher throughput, this is challenging in real-world deployments.

First, if requests with different TPOT values are scheduled on the same server, the server must operate at the smaller TPOT, significantly impacting the batch size. As shown in \fig\ref{fig:bdc_vs_tpot} and \fig\ref{fig:batch_size_vs_tpot}, when KV cache size is not the main bottleneck, increasing TPOT leads to a near-linear increase in maximum batch size. For example, for \pd{} with \((p,d) = (1000,4000)\), operating at a 40~ms TPOT can form a batch of approximately 150. If co-scheduled with a 20~ms TPOT request, the batch size drops to around 50, resulting in a near 1.5× cost increase.

Moreover, trivially assigning one cluster per TPOT tier also introduces inefficiencies. Since the request rate for each tier can vary, auto-scaling must be performed per SLO tier. However, auto-scaling is costly. When a new server is added, it takes significant time to schedule enough requests to reach the maximum batch size. Similarly, when removing a server, the system must wait for all requests to finish, resulting in underutilized cycles operating below the optimal batch size.

Finally, operating near the maximum batch size requires careful handling of tail latency. Since all requests on the server now have short SLOs, none can be terminated without violating their SLO. Therefore, the system cannot accommodate tail-latency-sensitive requests by terminating others. Tail latency must thus be proactively mitigated during scheduling.

\section{Design}

\subsection{\sys{} Overview}
In order to address the challenges and achieve high throughput multi-SLO serving, \sys{} first categorizes requests based on the TPOT , and then exploits fine-grained auto-scaling to manage the cluster size for each SLO by building a load gradient. For higher cluster utilization, \sys{} allows looser SLO requests to occupy the tighter SLO cluster when the looser SLO is full. To ensure high SLO attainment, \sys{} uses profile-based batch formation and wait time aware scheduling. Finally, \sys{} handles the TTFT through dynamic chunking and continuous chunked prefill prediction. We illustrate the overall design in Figure \ref{fig:slo_main}.

\subsection{Request Binning}

As illustrated in Sec \ref{sec:analysis}, mixing requests with different TPOT will result in suboptimal performance. Therefore, \sys{} maintains separate queues for different TPOT requests. Specifically, for co-location, since the KV cache transfer is not allowed between servers, \sys{} schedules both the prefill and decode phases of a request to its corresponding cluster. For disaggregation, \sys{} schedules the prefill phase across different TPOT requests while prioritizing the request with the nearest deadline, i.e., arrival time + TTFT. For the decode phase, \sys{} again schedules requests to the decode cluster with the corresponding SLO.

\subsection{Auto-scaling}

Request binning requires auto-scaling to allocate servers to different SLO tiers to handle load variation.
\sys{} will return a latency-sensitive server to the best-effort pool when the cluster is underutilized. When requests start pending for one SLO tier and the idle pool is not empty, \sys{} will assign the serving instance to the SLO.

Note that since the best-effort pool is also running the same model, joining a particular SLO tier simply requires dropping all requests and reconfiguring the batch size. Therefore, \sys{} assumes the scaling happens instantly.

In contrast, removing a server is more complex. Unlike traditional stateless workloads, where a server can be marked as empty and removed at any time, removing one LLM serving instance is more complex. Due to the large KV cache, terminating a server requires recomputing the on-the-fly request or migrating them to other servers. Only empty servers or prefill servers in PD-disaggregation can terminate at any time.

To facilitate auto-scaling, \sys{} prefers to schedule requests to the highest-load machine and try to maintain a load gradient inside each SLO tier. \sys{} will periodically check the requests in the last serving instance. If no requests are running on the server, the server will be returned to the best-effort pool.

\subsection{Lazy Promotion}

However, this auto-scaling strategy will cause the last server to be underutilized. To reduce the frequency of auto-scaling while increasing system utilization, \sys{} allows requests to be scheduled to clusters with tighter SLO, if and only if the current cluster is full. We name this lazy promotion.

Compared to eager promotion, which prioritizes scheduling requests to the tighter SLO server, lazy promotion slightly lowers the utilization of higher-tier servers while offering system-wide advantages.

1) When the lower tier is full and attempts to get a new server, both lazy and eager promotion behave similarly.

2) When the lower tier is not attempting to scale up or down, eager promotion may cause the tighter SLO cluster to scale up, as now more resources are dedicated to serving the lower-tier requests, and the tighter SLO requests cannot utilize the lower-tier’s idle server. Therefore, lazy promotion is beneficial.

3) Finally, only when the lower-tier is about to scale down does eager promotion create more opportunities, as it is more likely to find an empty server, which surpasses lazy promotion. However, having one more machine at the tighter SLO (case 2) is worse than having one more machine at the lower SLO (case 3). Thus, lazy promotion is better than eager promotion.

The presence of lower-tier requests will keep the last server in each cluster busy, which prevents the server from being removed. Specifically, even if the server is completely running requests at the lower tier, it will not be removed. To mitigate this, \sys{} modifies the auto-scaling criteria to put the server into a pending list when no request from the current tier is occupying the machine. In this state, the server can join the cluster serving the corresponding SLO as per the request inside the instance if that tier wants to scale up. Otherwise, it will wait until all requests finish and join the best-effort tier.

\subsection{Profile-Based Batch Formation}

While building and maintaining a load gradient facilitates auto-scaling, it poses more challenges to the router, as requests face a higher chance of SLO violation. The key to achieving a high load while maintaining the SLO is accurate iteration time prediction.

Firstly, \sys{} predicts the iteration time of serving an instance based on profiling. As illustrated in Section \ref{sec:bg:inference}, the iteration time is the sum of GEMMs and collective communications, which depend on the batch size, decode attention, which depends on the KV cache size, and prefill attention, which depends on the prefill length. For common workloads with sequence lengths less than 10K, the prefill time, when considering chunked prefill, is significantly less than the other parts. Therefore, \sys{} only considers batch size and KV cache size. Through profiling, \sys{} builds a map of (batch size, KV cache size) to execution time.

Moreover, predicting the length of the request is both costly and inaccurate. Indeed, \sys{} simplifies the problem by just predicting the output length using the average decode length. Due to the deadline-based SLO, the strict TPOT constraint is loosened. Therefore, while some iterations, due to misprediction of the output length, may surpass the TPOT, other less-occupied cycles can compensate for the delay, which can still meet the DSLO.

When routing requests, \sys{} iterates through the servers and estimates the iteration time after admitting the new requests. Specifically, \sys{} simulates future iterations and computes the maximum KV cache size as requests grow in length, and exits when finished. Based on the largest KV cache size and current token batch size, \sys{} uses the profiling table to admit the request when the predicted iteration time is less than the TPOT.

\subsection{Wait-Time-Aware Scheduling}

While profile-based batch formation can maintain good TPOT attainment, it is only effective from the second decode token (i.e., third generated token). The first token, regulated by the TTFT, and the second token, regulated by TTFT + TPOT, incur queueing time. The queueing time again can be broken down into time in the pending queue (pending time) and, after being scheduled to a particular server, waiting for the server to finish the current iteration (wait time). In the worst case, when scheduling a request that generates the first token exactly at TTFT for PD-disaggregation, even if all servers are low in load and it is the only pending decoding request, it must be scheduled to the server that finishes the current cycle very soon. Otherwise, the wait time and iteration time will cause the second token to miss the deadline. Therefore, when judging whether a server can admit a request, \sys{} considers the wait time along with iteration time to find the highest-load but SLO-achievable server.

\subsection{Handling TTFT}

\textbf{PD-disaggregation: dynamic chunking}
Attaining TTFT for PD-disaggregation is more straightforward, as no decode request will interfere with the prefill. \sys{} schedules requests based on the deadline and estimated prefill speed, and assigns the request to the highest-load server that can still achieve TTFT. However, while TTFT can allow a very large batch size, always scheduling requests until that limit is not feasible, as the wait time of the next round would be significant, thus increasing the chance of violation for the next round. Therefore, \sys{} still limits the token batch size similar to chunked prefill. However, chunked prefill has tail latency issues. Specifically, if the token budget is 1024, whose iteration time is $T$, and the request sequence length is 2050. If not using chunked prefill, the time to finish the prefill is around $2T$. However, if using chunked prefill, the request needs 3 cycles to finish, which is $3T$. In the worst case, if the first iteration is already filled with 1023 tokens, the request needs 4 cycles, i.e., 1+1024+1024+1 to finish the prefill, which is $4T$. These factors can greatly affect TTFT attainment. \sys{} instead uses a dynamic chunked prefill to allow some flexibility in the batch size. Specifically, when the remaining length is less than 2 times the budget, \sys{} will handle the remaining tokens at once, while not admitting new requests to fill that 2 times the budget. This allows almost every request to reduce one iteration to finish while still controlling the wait time. Moreover, \sys{} considers the additional cycle due to the server already having some requests scheduled and reroutes to other machines if \sys{} predicts a TTFT violation.

\textbf{Co-location: continuous chunked prefill prediction}
Ensuring TTFT attainment for co-location is more challenging due to the presence of decode requests. Specifically, to attain TPOT, the prefill is chunked into more pieces. \sys{} estimates the completion time using iteration time and the number of chunks predicted at the time of scheduling. However, as the KV cache grows, if the same chunk size is maintained until the prefill is completed, some iterations may violate TPOT. Therefore, \sys{} takes a more conservative approach. \sys{} only admits requests if the predicted chunk size can be maintained throughout the prefill process. Otherwise, \sys{} will look for other lower-load machines.

\section{Evaluation}
\label{sec:evaluation}
\begin{table*}[t]
\centering
\caption{Trace Input/Output Percentile Statistics}
\begin{tabular}{lcccccccccccc}
\toprule
\textbf{Trace File} & \multicolumn{6}{c}{\textbf{Input}} & \multicolumn{6}{c}{\textbf{Output}} \\
\cmidrule(lr){2-7} \cmidrule(lr){8-13}
& 25\% & 50\% & 75\% & 90\% & 95\% & 99\%
& 25\% & 50\% & 75\% & 90\% & 95\% & 99\% \\
\midrule
\texttt{uniform\_4096\_1024} & 2047 & 4093 & 6149 & 7377 & 7785 & 8108 & 510 & 1023 & 1535 & 1843 & 1944 & 2027 \\
\texttt{uniform\_512\_512} & 255 & 511 & 768 & 921 & 973 & 1013 & 256 & 511 & 768 & 922 & 973 & 1014 \\
\texttt{mooncake\_conversation} & 2320 & 6923 & 15400 & 27571 & 39583 & 85401 & 159 & 350 & 472 & 597 & 698 & 1136 \\
\texttt{mooncake\_synthetic} & 277 & 11587 & 23286 & 38737 & 49009 & 66458 & 10 & 68 & 250 & 390 & 522 & 768 \\
\texttt{mooncake\_toolagent} & 3228 & 6346 & 7468 & 16818 & 26175 & 61824 & 12 & 30 & 355 & 506 & 600 & 890 \\
\texttt{lmsys} & 12 & 28 & 82 & 301 & 430 & 750 & 39 & 140 & 338 & 512 & 519 & 853 \\
\texttt{sharegpt} & 16 & 36 & 158 & 818 & 1613 & 3421 & 131 & 280 & 445 & 682 & 846 & 1001 \\
\texttt{splitwise} & 396 & 1019 & 1186 & 2735 & 4083 & 4142 & 85 & 130 & 395 & 425 & 451 & 601 \\
\bottomrule
\label{tab:trace_stat}
\end{tabular}
\end{table*}
\subsection{Evaluation Setup}

\textbf{Evaluation Methods.}
We implement \sys{} in C++ for high performance. Since multi-SLO serving requires large-scale deployment (tens to hundreds of instances) to emulate realistic production environments, real-world evaluation becomes costly and challenging. Additionally, at the start of the service, all servers are idle, making metrics such as Time-To-First-Token (TTFT) and Tokens-Per-Output-Time (TPOT) meaningless. Thus, a significant warm-up period (tens of minutes) is required to collect meaningful performance data. Considering the large number of configuration combinations (hundreds), evaluating every setting on physical hardware is impractical.

Due to these constraints, we evaluate \sys{} via simulation. The simulator uses real kernel-level profiling data from vLLM~\cite{vllm} to reflect actual per-stage cycle times. It runs at a 1ms simulation timestep to balance precision and performance, as TPOT values typically fall within tens of milliseconds.

\textbf{Model and GPU.}
We use LLaMA3.1-8B~\cite{llama3} as the target model, as it supports Grouped-Query Attention (GQA), representative of both the LLaMA and Qwen~\cite{qwen} model families. We collect profiling traces using an NVIDIA H200 GPU, a widely adopted inference accelerator in production environments.

\textbf{Datasets.}
We use a combination of synthetic data and real-world traces to cover diverse request characteristics, including mooncake trace~\cite{qin2024mooncakekvcachecentricdisaggregatedarchitecture}, lmsys~\cite{zheng2023lmsyschat1m}, sharegpt~\cite{sharegpt}, splitwise~\cite{splitwise}. Table~\ref{tab:trace_stat} summarizes the p25, p50, p75, p90, and p99 of input/output lengths for each dataset. To evaluate steady-state performance, we randomly sample 300,000 requests per dataset.

\textbf{SLO Tiers.}
Since existing traces do not include SLO annotations per request, we empirically define them. TTFT is sampled uniformly from three values: 300ms, 500ms, and 1000ms.  TPOT tiers are set to 20ms, 30ms, 50ms, and 100ms, with probabilities 10\%, 20\%, 30\%, and 40\%, respectively. Note the minimal per-token latency of vLLM on H200 ( batch size = 1 and context length = 1) is around 15ms, and 100ms is enough for saturating the engine performance.
Additionally, each request is only assigned an SLO if it is achievable assuming immediate dispatch to an idle server.

\textbf{Scheduling Policies.}
We evaluate the following scheduling policies:
\begin{itemize}
    \item \textbf{\pdsys{}}: \sys{} deployed on prefill-decode disaggregated systems.
    \item \textbf{\cosys{}}: \sys{} deployed on co-located systems.
    \item \textbf{PD-Random}: Randomly choosing servers on disaggregated systems
    \item \textbf{CO-Random}: Randomly choosing servers on co-located systems.
    \item \textbf{PD-Minimal}: Assigning requests to the lowest cycle-time server in a disaggregated setup.
    \item \textbf{CO-Minimal}: Assigning requests to the lowest cycle-time server in a co-located setup.
    \item \textbf{CO-Chunk}: A chunked scheduler in a co-located setup, statically configured with a maximum token budget. In our evaluation, we iterate over different token budgets and select the one yielding either the highest SLO attainment or lowest number of servers needed for a given request rate.
\end{itemize}

\begin{figure*}[p]
    \centering
    \includegraphics[height=0.12\textheight]{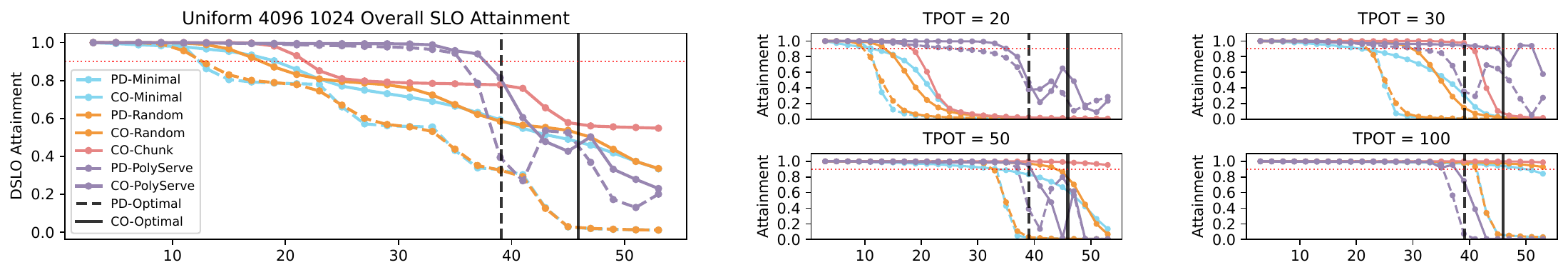}
    \includegraphics[height=0.12\textheight]{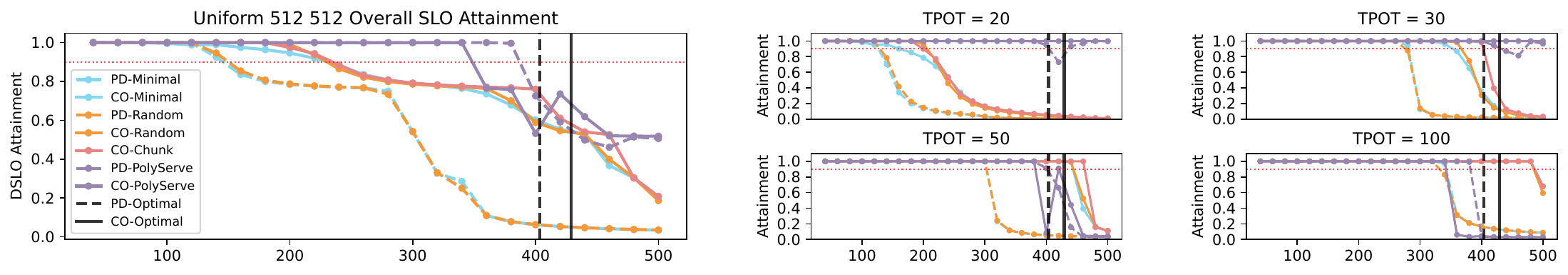}
    \includegraphics[height=0.12\textheight]{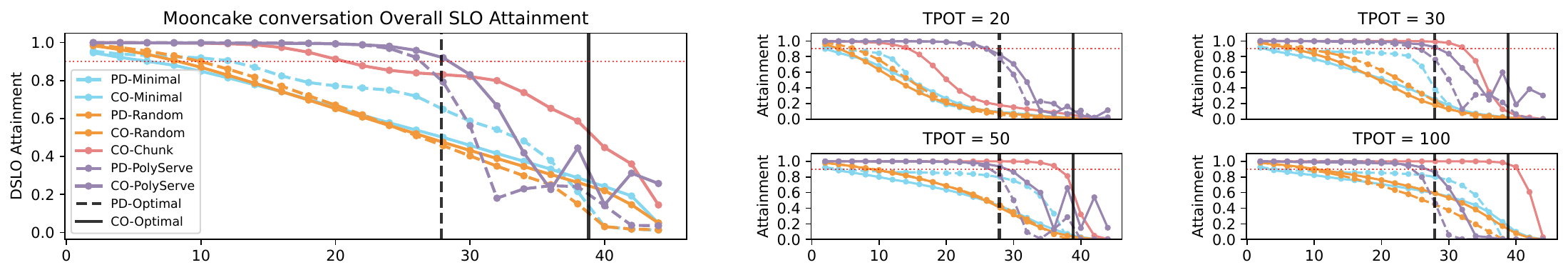}
    \includegraphics[height=0.12\textheight]{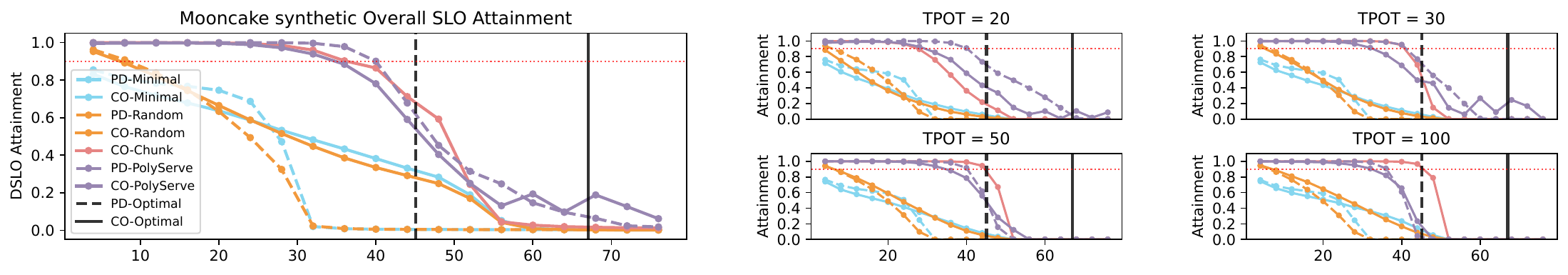}
    \includegraphics[height=0.12\textheight]{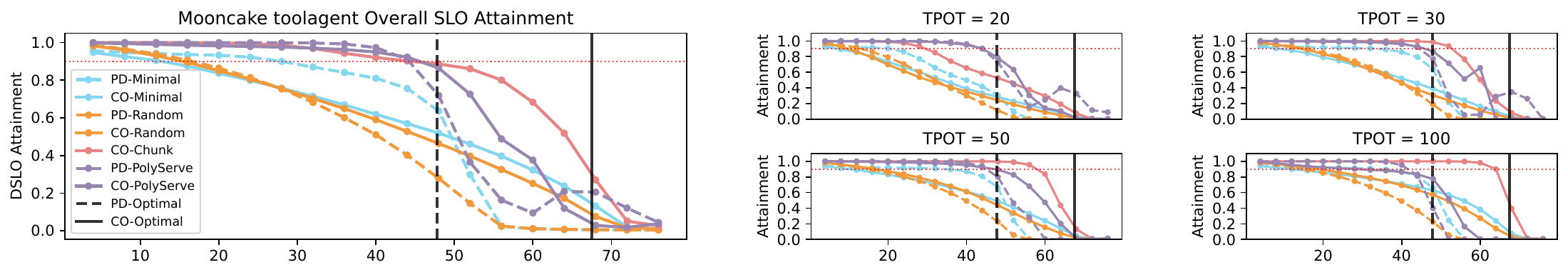}
    \includegraphics[height=0.12\textheight]{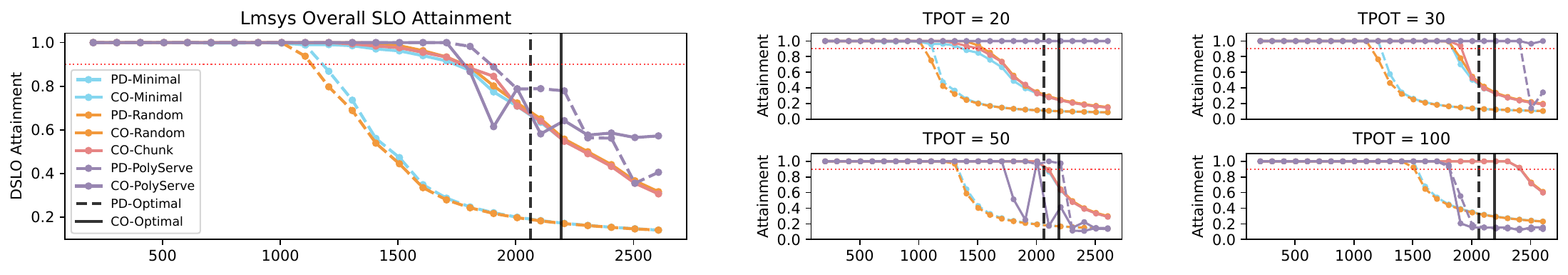}
    \includegraphics[height=0.12\textheight]{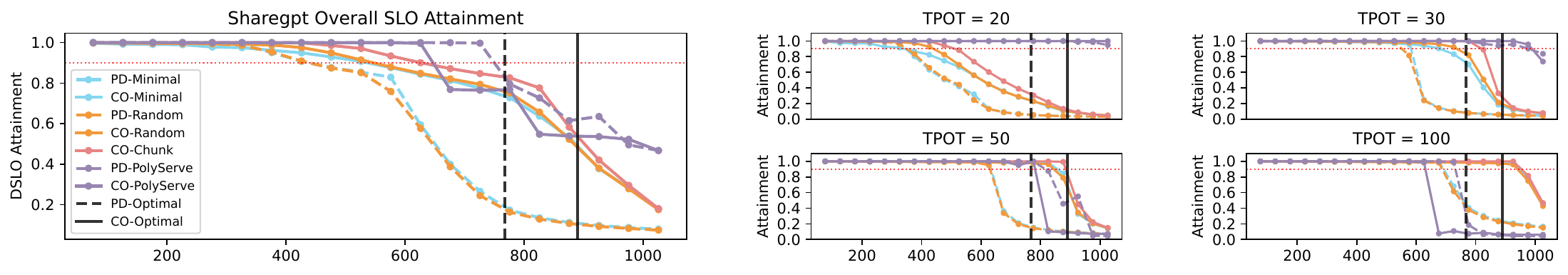}
    \includegraphics[height=0.12\textheight]{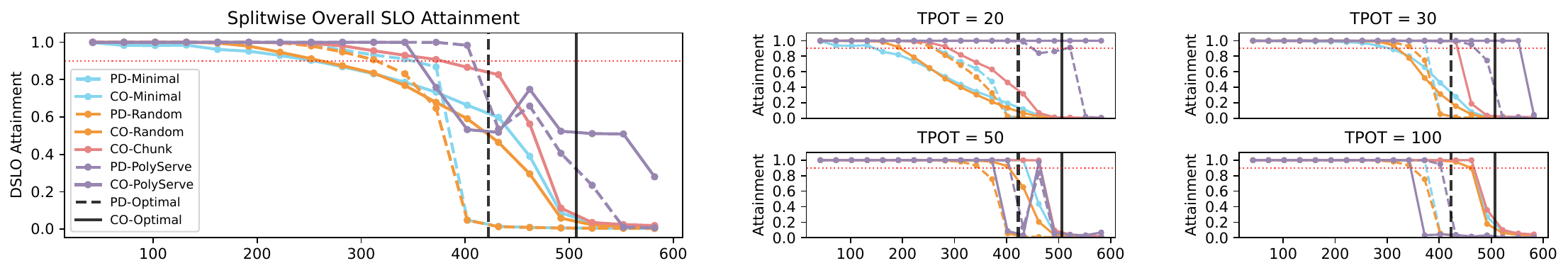}
    \caption{Overall DSLO attainment and DSLO attainment breakdowns at various request rates.}
    \label{fig:eval_goodput}
\end{figure*}

\subsection{Goodput Evaluation}

We set the number of serving instances to 20 to mimic a small-scale service provider. The request rate for each dataset is varied from 20\% to 120\% of the optimal throughput. The Requests arrive according to a Poisson process.

Each row of the figure reports the overall SLO attainment, as well as breakdowns by TPOT tier. \pdsys{} and \cosys{} consistently achieve high SLO attainment across the load spectrum. At 90\% attainment, \pdsys{} and \cosys{} provide on average \datagoodputgainpd and \datagoodputgainco more goodput than the best baseline. Note that, as shown in the TPOT breakdown, \pdsys{} and \cosys{} achieve SLO near-uniformly for all different tiers, while other baselines reduce SLO attainment, especially for low TPOT requests. 

The performance of \pdsys{} and \cosys{} highly depend on the input and output statistics of the trace. Compared to \cosys{}, \pdsys{}'s goodput lies between $-7.4\%$ to $+15.9\%$. Generally, for long traces, the \cosys{} have higher throughput, thus have higher goodput. For shorter traces, \pdsys{} is easier to manage and thus have goodput benefits. 

Overall, across all traces, \pdsys{} achieves \datagoodputpdopt of the optimal prefill-decode disaggregate goodput, while \cosys{} achieves \datagoodputcoopt of the optimal co-location goodput. 

\begin{figure}[t]
    \centering
    \includegraphics[width=\columnwidth]{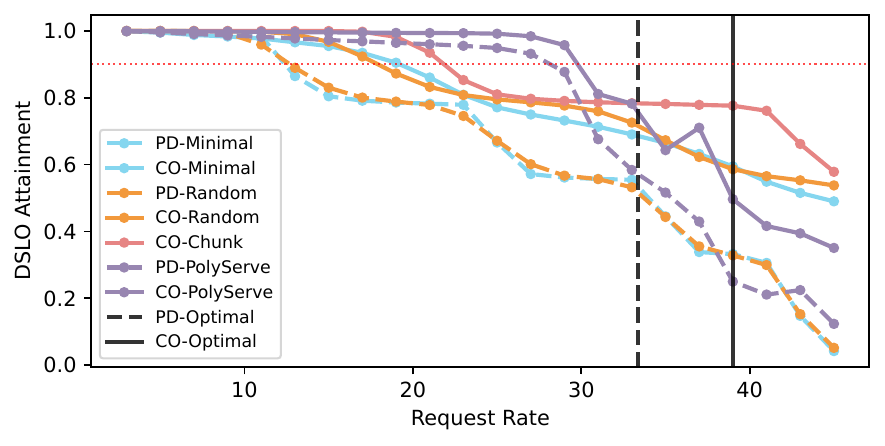}
    \caption{DSLO attainment when changing request distribution across time. \pdsys{} and \cosys{} can gracefully handle the request rate variation and achieve high goodput}
    \label{fig:eval_burst}
\end{figure}

\subsection{Burstiness Evaluation}

Building on the goodput setup, we uniformly sample input lengths from [1, 8192] and output lengths from [1, 2048]. The first 150,000 requests follow a TPOT distribution of 10\% at 20ms, 20\% at 30ms, 30\% at 50ms, and 40\% at 100ms. For the remaining 150,000 requests, the distribution is inverted: 40\% at 20ms, 30\% at 30ms, 20\% at 50ms, and 10\% at 100ms. All requests follow a Poisson arrival process.

As shown in Figure~\ref{fig:eval_burst}, \pdsys{} and \cosys{} leverage internal autoscaling to absorb bursts in high-priority requests, achieving \databurstgainpd and \databurstgainco goodput at 90\% attainment compared to the best baseline.

\begin{figure}[t]
    \centering
    \includegraphics[width=\columnwidth]{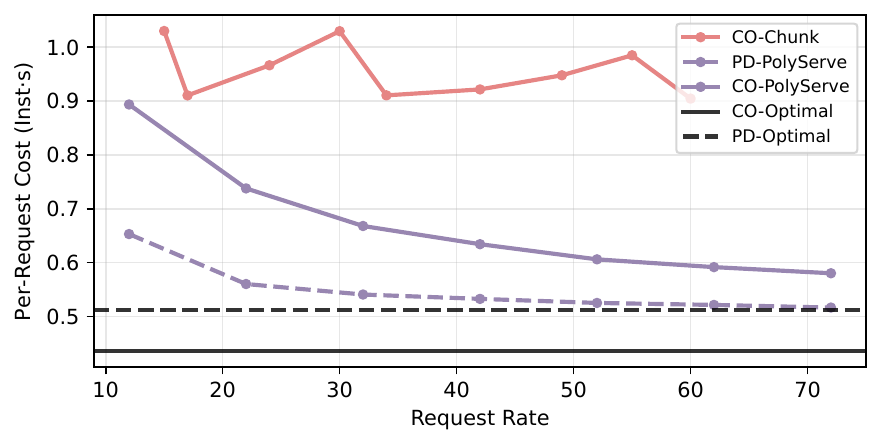}
    \caption{The per-request serving cost for various algorithms and request rates when given enough instances. }
    \label{fig:eval_cost}
\end{figure}
\subsection{Cost Evaluation}

In this experiment, we assume an unlimited server pool and report total cost measured as $instances \times seconds$. For baseline, we select the best performing CO-Chunk and vary the number of servers to just achieve 90\% SLO attainment, while for our \pdsys{} and \cosys{}, we simply provide enough servers and let auto-scaling to determine the usage. This evaluation is challenging as a good policy must use as few servers as possible while attaining the SLO. 

Figure~\ref{fig:eval_cost} shows the cost per request across increasing request rates, where all policies meet 90\% SLO attainment. \pdsys{} and \cosys{} consistently achieve lower cost. 

\subsection{Sensitivity Study on Server Nums}

We evaluate the robustness of \sys{} by varying the number of servers and the number of SLO tiers.

\begin{figure}[t]
    \centering
    \includegraphics[width=\columnwidth]{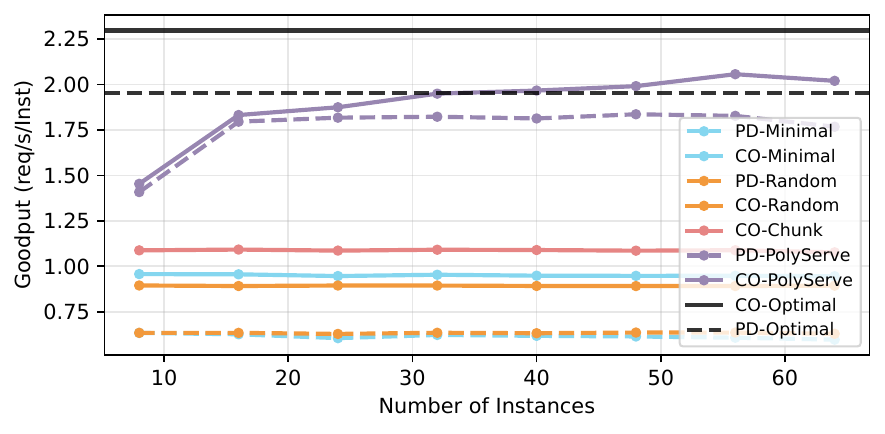}
    \caption{Per-instance goodput when changing the number of instances. As the number of instances increases, the per-instance goodput also increases until saturation.  }
    \label{fig:eval_num_server}
\end{figure}

In Figure~\ref{fig:eval_num_server}, we vary the number of servers from 8 to 64 (step size: 8), with uniform-4094-1024 trace, and compare goodput per instance under each policy. \pdsys{} and \cosys{} consistently outperform all baselines, and as the number of servers increases, the per-instance throughput increases due to reduced fragmentation.

\subsection{Scheduler Efficiency Evaluation}
To assess scheduling throughput, we measure the maximum number of requests the scheduler can arrange per second as the number of servers increases. Based on profiling, a single server can handle $4825$ requests per second, supporting over $100$ servers in real-time scheduling without performance degradation, where the performance already saturates. \sys{} can further scale by introducing more schedulers that manage independent servers.

\section{Related Work}

\subsection{LLM Serving Frameworks}

Achieving higher throughput is a critical goal for LLM serving frameworks. Existing systems like vLLM~\cite{vllm} use paged attention to reduce memory fragmentation and support larger batch sizes. SGLang~\cite{zheng2024sglangefficientexecutionstructured} implements radix attention to enable prefix sharing. NanoFlow~\cite{zhu2025nanoflowoptimallargelanguage} overlaps operations with different resource requirements to push throughput at high batch sizes. In contrast, \sys{} maintains all queues in the router, easing the burden on serving frameworks and dispatching requests to the actual serving engine. As a result, \sys{} requires minimal changes to integrate with different frameworks and is orthogonal to engine-level optimizations.

\subsection{SLO-Aware Optimizations}

SLO-attainment optimization has also gained significant attention. For example, SarathiServe\cite{sarathi-serve} proposes chunked prefill, which reduces interference between prefill and decode phases. DistServe\cite{distserve} separates servers into prefill and decode clusters to isolate performance. While debates continue on which strategy is more suitable for production deployments, we provide support for both cases and show that the performance differences are not significant; the choice often depends on the infrastructure setup.

Several works further optimize beyond prefill-decode disaggregation and chunked prefill. For example, MoonCake~\cite{qin2024mooncakekvcachecentricdisaggregatedarchitecture} proposes a KV-cache-centric architecture that gracefully handles overload and achieves higher throughput. Tempo\cite{zhang2025tempoapplicationawarellmserving} focus on improve the SLO attainment given inaccurate infomation about the output length. Niyama \cite{goel2025niyamabreakingsilos} repurposes the slack in TTFT to identify additional batching opportunities and improve goodput. SLOs-Serve \cite{chen2025slosserveoptimizedservingmultislo} identifies the set of SLO-achievable requests and uses dynamic programming to optimize token allocations. However, most of these works focus on local settings with diverse request lengths. In contrast, \sys{} is designed for large-scale deployment, where separating requests by SLO type is feasible and advantageous.

\subsection{Large-Scale Deployment}

Existing research also explores large-scale deployment of LLMs. AIBrix \cite{theaibrixteam2025aibrixscalablecosteffectivelarge} develops a control plane to manage serving engines at scale, handling auto-scaling, KV-cache management, and basic request routing. ServerlessLLM \cite{serverlessyaofu} investigates fast checkpoint loading to reduce autoscaling startup latency. $\lambda$Scale~\cite{yu2025lambdascaleenablingfastscaling} loads models during execution to further reduce startup times. These works address cluster-level auto-scaling and management, which are foundational to \sys{}. However, since \sys{} schedules requests with different TPOTs separately, it demands more fine-grained auto-scaling to balance both SLO attainment and system efficiency.

\section{Conclusion}
In this work, we demonstrate the benefit of providing SLO-tiered pricing and present the challenges to serving requests with multiple SLOs at scale. We propose \sys{}, a scheduling policy for large-scale multi-SLO workloads. Specifically, \sys{} group requests by TPOT and assigns a different cluster for different SLO tiers. \sys{} adjust the size of each cluster through fine-grained auto-scaling and manage the tail latency by wait-time aware scheduling, dynamic chunking, and continuous chunked prefill prediction. \sys{} achieves \datagoodputgainpd{} and \datagoodputgainco{} goodput gain compared to existing policies for prefill-decode disaggregation and co-location, respectively.

\bibliographystyle{plain}
\bibliography{_reference}

\end{document}